\documentclass[12pt]{article}

\usepackage{rotating}
\usepackage{epsfig}

\newcommand{\re}[1]{\ (\ref{#1})}
\newcommand{\nn}{\nonumber}
\newcommand{\ed}{\end{document}}
\newcommand{\be}{\begin{equation}}
\newcommand{\ee}{\end{equation}}
\newcommand{\ba}{\begin{eqnarray}}
\newcommand{\ea}{\end{eqnarray}}
\newcommand{\baz}{\begin{eqnarray*}}
\newcommand{\eaz}{\end{eqnarray*}}
\newcommand{\bb}{\begin{thebibliography}}
\newcommand{\eb}{\end{thebibliography}}
\newcommand{\ct}[1]{${\cite{#1}}$}

\newcommand{\bi}[1]{\bibitem{#1}}
\begin{document}
\begin{center}
\thispagestyle{empty} \vskip 3cm {\LARGE \bf The Pauli form factor
of the quark induced by
 instantons}
\vskip 0.5cm
N.I. Kochelev$^{1,2}$\\
\vspace{5mm}
{\small\it
$^1$ BLTP,
Joint Institute for Nuclear Research,\\
Dubna, Moscow region, 141980 Russia\\
$^2$ Institute of Physics and Technology, Almaty, 480082, Kazakhstan}

\vspace*{1cm}
\end{center}
\begin{abstract}
\noindent The non-perturbative contribution to the Pauli form factor
of the quark, $F_2(Q^2)$, is calculated within an instanton model
for the QCD vacuum. It is shown that the instantons give a large
negative contribution to the form factor.

\end{abstract}
\newpage

The interaction of photons with hadrons is one of the most
powerful tools to investigate the structure of strong
interactions. The electromagnetic form factors of hadrons are
 now widely discussed\cite{reviews}. This increased
interest in the electromagnetic form factors is related to the
significant theoretical progress in understanding the connection
between form factors and structure functions of hadrons through
the generalized parton distributions (GPD) \cite{gpd}, and to 
the need to explain the new experimental data from Jefferson Lab
\cite{JLab} for the nucleon and pion form factors at large $Q^2$.
One of the basic components of the calculation of the form factors
within the constituent quark model are the electromagnetic form
factors of quarks. It has been shown that the non elementarity of
the constituent quarks plays a crucial role for the understanding
of the parton distributions and form factors
\cite{altarelli,scopetta}. Moreover quarks form factors have been
instrumental in the investigation of the scaling violations in
deep-inelastic scattering \cite{ricco}. We should also mention
that in order to explain the high twist effects in Drell-Yan
processes \cite{dy}, various perturbative and non-perturbative
gluonic corrections to the photon-quark vertex must be taken into
account \ct{brodsky}.

The general formula for the photon-quark vertex is \be
\Gamma_\mu=F_1(Q^2)\gamma_\mu+\frac{iq_\nu\sigma_{\mu\nu}}{2m_q}F_2(Q^2),
\label{ff} \ee where $m_q$ is the mass of quark,
$\sigma_{\mu\nu}=i(\gamma_\mu\gamma_\nu-\gamma_\nu\gamma_\mu)/2$
and $F_1(Q^2)$, $ F_2(Q^2)$  are the Dirac and Pauli form factors,
correspondingly. There exist calculations, both perturbative and
non-perturbative, of QCD contributions to the $F_1(Q^2)$ form
factor (see \cite{ermolaev}, \cite{renormalon} and references
therein).

The usual way to incorporate the non-perturbative QCD dynamics
into the calculation of quark form factors is the consideration of
the renormalon contributions \ct{renormalon}. However, renormalon
contributions amount only to a part of the QCD non-perturbative
effects.

The instanton model for the QCD vacuum is now one of the most
successful models for the description of non-perturbative effects
in strong interactions (see the review \cite{shuryak}). Within
this model, the QCD vacuum contains  complex topological
structures related to existence of the instantons, strong
fluctuations of the gluon vacuum fields. In QCD the instantons
play an important role in describing the realization of chiral
symmetry breaking and the origin of the masses of constituent
quarks and hadrons. In refs. \ct{kochdor}-\ct{diakonov} it was
shown that the instantons produce also specific non-perturbative
effects in various hadronic reactions. Recently, the first step in
the calculation of instanton contributions to the quark form
factor was done in refs. \ct{dorokhov}. Within the Wilson integral
approach, it was shown that the instantons lead to a finite
renormalization of the next-to-leading perturbative correction to
the $F_1(Q^2) $ form factor.

One of the most important features of the instanton induced
quark-quark and quark-gluon interactions, in comparison with the
perturbative gluon exchange, is the possibility of having quark
chirality-flip due to the instanton field. From this point of
view, the chirality-flip $F_2(Q^2)$ quark form factor plays the
role of a filter of instantons, since the perturbative
contribution to this form factor at large $Q^2>>m_q^2$, is small,
i.e.  proportional to $m_q^2/Q^2$ \ct{ermolaev2}. Therefore, the
dominance of the instanton contribution to $F_2(Q^2)$ for the
light $u-, d-, s- $ quarks is expected even at large  momentum
transfers.

At small $Q^2$, the Pauli form factor receives also contributions
from both, perturbative gluon exchanges and  non-perturbative QCD
effects \ct{elias}. It is well known that the perturbative
Schwinger correction \ct{schwinger} (Fig.1a) to the quark
anomalous magnetic moment $\kappa_q=F_2(0)$ is rather large and
{\it positive} \be
\kappa_q^{pQCD}=\frac{2\alpha_s(m_q^2)}{3\pi}\approx 0.14
\label{pQCD} \ee
where the so-called analytic running constant \cite{shirkov}
\be
 \alpha_s(Q^2)=\frac{4\pi}{\beta_0}\left.[\frac{1}{log(Q^2/\Lambda^2)}+
\frac{\Lambda^2}{\Lambda^2-Q^2}\right.] \label{shirkov1} \ee
has been used with $\Lambda\approx 300 MeV$ and we have used
$m_q\approx 350 MeV$ for the constituent mass of the quark in the
hadron, which corresponds to the magnitude of the quark effective
mass in the instanton liquid model \ct{diakonov1}. Such large
contribution would result in a modification of the constituent
quark predictions for the electromagnetic properties of hadrons,
which, however, describe rather well the data without a sizeable
quark anomalous magnetic moment (see discussion in \cite{elias}).
Furthermore, from the analysis of the scaling violations observed
in deep-inelastic scattering at JLab, performed within the
constituent quark model \cite{ricco},  a {\it negative} value of
$F_2(Q^2)$ form factor at low $Q^2$ has been obtained. It is
therefore very difficult to explain the behaviour of $F_2(Q^2)$
using only  the perturbative QCD framework.

In this Letter we calculate the instanton contribution to the
$F_2(Q^2)$ form factor of the quark (Fig.1b).

\begin{figure}[htb]
\hspace*{2cm}\mbox{\begin{turn}{-90}
\epsfig{file=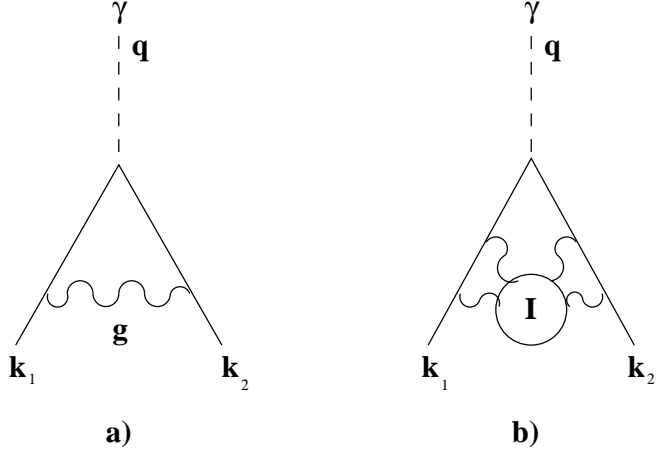,width=6cm}
\end{turn}}
\vskip 0.5cm
\caption{\it  The contribution  to the  quark form factor $F_2(Q^2)$
from a) one gluon  exchange and b)
instanton}
\end{figure}
It is convenient to use the two component Dirac spinors
\begin{equation}
\Psi={\chi_L\choose{\chi_R}},
\nonumber
\end{equation}
and the following representation for $\gamma$ matrices:
\begin{equation}
\gamma_\mu={{0\ \  \sigma_\mu}\choose{\bar\sigma_\mu \ \  0}},
\nonumber
\end{equation}
where, in Minkowski space-time, $\sigma_\mu=(1, \vec{\sigma_\mu})$,
$\bar\sigma_\mu=(1,- \vec{\sigma_\mu})$, and in Euclidean space-time
 $\sigma_\mu=(-i\vec{\sigma_\mu},1)$,
 $\sigma_\mu=(i\vec{\sigma_\mu},1)$.
The photon-quark vertex which corresponds to the transition of the
incoming left-hand quark to the outgoing right-hand quark in
Fig.1b, is \be A_\mu^{LR}
=F_2(Q^2)\frac{q_\nu\chi_R^+(\sigma^\nu\bar\sigma^\mu
-\sigma^\mu\bar\sigma^\nu)\chi_L}{4m_q}, \label{form1} \ee
As was discovered  by t'Hooft \ct{thooft}  there is a zero mode
for the quark in the instanton field. This mode has the chirality
required to produce a helicity flip of scattered quark. The
non-zero contribution to $A_\mu^{LR}$ arises when either the
incoming, or the outgoing quarks in Fig.1b scatter through the
zero-mode of the instanton field. The result of calculation is
\ba
A_\mu^{LR,inst}&=& -i\int dUd\rho\frac{d(\rho)}{\rho^5}\chi^+(k_2)
(ik_2)[ \Phi_0(-k_2)V_\mu(q,k_1) \nn\\ &+&\bar
V_\mu(q,-k_2)\Phi_0^+(ik_1)](ik_1) \chi_L(k_1)/m_q,
\label{formula1} \ea where \ba V_{\mu}(q,k_1)&=&\int
d^4xe^{-iqx}\Phi(x)^+\bar\sigma_\mu
 S_{nz}^{(I)}(x,k_1)\nn\\
\bar V_\mu(q,-k_2)&=&\int d^4xe^{-iqx}\bar S_{nz}^{(I)}(-k_2,x)
\sigma_\mu\Phi(x), \label{not} \ea
where $\Phi_0 (\Phi_0^+)$ and $S_{nz}^{(I)}$, $\bar S_{nz}^{(I)}$
are the zero-mode and non-zero mode quark propagators for the
incoming (outgoing) quarks. In Eq.(\ref{formula1}), $\rho$ is the
instanton size, $d(\rho)$ is the instanton density,  $dU$ stands
for the integration over instanton orientation in colour space,
and $k\equiv k_\mu\sigma^\mu$, $\bar k\equiv k_\mu\bar\sigma^\mu$
for any four-vector $k_\mu$ \ct{ringwald1}. In Eq.(\ref{formula1})
the continuation from Euclidean space-time to Minkowski space-time
should be done by substituting $ k_{i4}\rightarrow
(-i+\epsilon)\sqrt{\vec{k}_i^2},\
  q_4\rightarrow (-i+\epsilon)q_0 $.

For the light quarks $k_{1,2}^2=m_q^2\approx 0$ the zero-mode is
well known asymptotically
\ba ((ik_2)\Phi_0(-k_2))^{\alpha
i}&\rightarrow& 2\pi\rho
 U^i_\beta\epsilon^{\beta\alpha}
\nn\\
(\Phi_0(k_1)^+(i\bar k_1))^{\alpha i}&\rightarrow & 2\pi\rho
\epsilon^{\beta\alpha}(U^+)_\beta^i,
\label{zero}
\ea
where $U$ is the orientation matrix of the instanton in colour
space. The asymptotic behaviour of the non-zero mode propagators
at $Q^2\gg m_q^2$ was found recently in ref. \ct{ringwald1}
\ba
S_{nz}^{(I)}(x,k_1)(ik_1)&\rightarrow&-\frac{e^{-ik_1x}}{\sqrt{\Pi(x)}}
\left(1+(1-e^{ik_1x})\frac{\rho^2}{x^2}\frac{Ux\bar k_1U}{2k_1x}\right)\nn\\
(ik_2)\bar S_{nz}^{(I)}(-k_2,x)&\rightarrow&-\frac{e^{ik_2x}}{\sqrt{\Pi(x)}}
\left(1+(1-e^{-ik_2x})\frac{\rho^2}{x^2}\frac{Uk_2\bar xU}{2k_2x}\right),
\label{propag}
\ea
where $\Pi(x)=1+\rho^2/x^2$.
By using these formulas, after integration over the colour orientations
(see \ct{shuryak}, \ct{ringwald1}) and a simple algebra, we get
\ba
A_\mu^{LR,\ inst}&=&
\frac{4\pi^2}{3Q^2m_q}\int d\rho\frac{d(\rho)}{\rho^3}
(\rho QK_1(\rho Q)-1)\nn\\
& &q_\nu\chi_R^+(\sigma^\nu\bar\sigma^\mu
-\sigma^\mu\bar\sigma^\nu)\chi_L. \label{ff2} \ea From \re{ff2}
and \re{form1}, the following result is obtained for the instanton
contribution to the  form factor: \be
F_2^{inst}(Q^2)=\frac{16\pi^2}{3Q^2}\int\frac{d\rho
d(\rho)}{\rho^3} (\rho QK_1(\rho Q)-1) \ . \label{final1} \ee
For a numerical estimate of this contribution, we use the
instanton liquid model suggested by Shuryak \ct{shuryak3}. Within
this model, the density of instantons is
\be
\frac{d(\rho)}{\rho^5}={n_I}\delta(\rho-\rho_c), \label{density}
\ee
where $n_I\approx 1/2 fm^{-4}$,  $\rho_c\approx 1/3 fm$  are
the average density and size of instantons in the QCD vacuum,
correspondingly. The final result for the form factor is
\be
F_2^{inst}(Q^2)=\frac{8f}{3}\frac{(\rho_cQK_1(\rho_cQ)-1)}{\rho_c^2Q^2},
\label{final} \ee
where $f=n_I\pi^2\rho_c^4\approx 0.1$ is the so-called packing
fraction (diluteness) of the instantons in the vacuum.
\begin{figure}[htb]
\hspace{3cm}\mbox{\epsfig{file=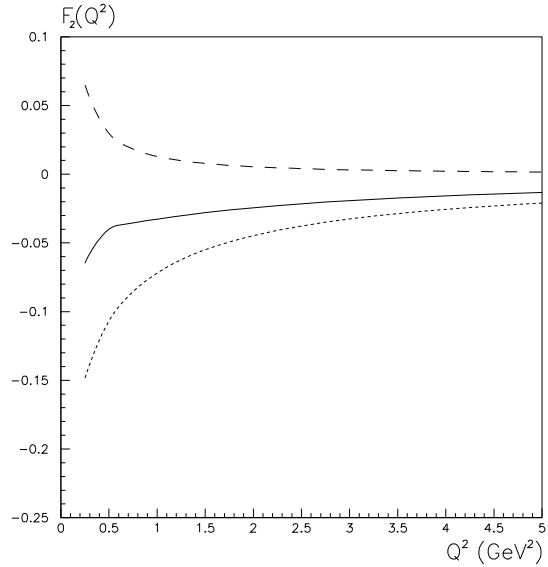,width=8cm}}
\caption{\it   The contributions to the  Pauli form factor of the
quark from a) instantons (short-dashed line), b) perturbative
gluon exchanges (long-dashed line) and c) their sum (solid line).}
\end{figure}
In the Fig.2 the contribution to the quark Pauli form factor
arising  from instantons \re{final} is compared with the
result of the double-logarithmic perturbative calculation \ct{ermolaev2}
\be
F_2(Q^2)^{pQCD} =
\frac{\alpha_s(Q^2)C_F}{\pi}\frac{m_q^2ln(Q^2/m_q^2)}{Q^2}
exp\left(-\frac{\alpha_s(Q^2)C_F}{4\pi}ln^2( \frac{Q^2}{m_q^2} )\right) \ ,
\label{pqcd}
\ee
at $Q^2=0.25\div 5 GeV^2 $ for the quark  mass\footnote{For the current mass of the quark
$m_q\approx 5 MeV$, the pQCD contribution to $F_2$ form factor is
very tiny ($10^{-5}\div 10^{-9})$ in this interval of $Q^2$.}  $m_q\approx 350 MeV$
and analytic running  coupling constant
$\alpha_s(Q^2)$ given in \re{shirkov1}.

It is evident that the instantons give a large negative
contribution to the quark form factor in the region of momentum
transfer of order of a few $GeV^2$. This contribution is larger
then the positive contribution coming from the perturbative
exchange, and one may expect that the total Pauli form factor of
 the quark should be negative.

Therefore the instanton corrections to the chirality-flip  quark-photon
vertex  can not be neglected in the photon induced
reactions in this kinematic region. The application of the obtained
result for the specific reactions will be the subject of forthcoming
publications.

\vspace{0.2cm}

The author is thankful to I.O. Cherednikov, A.E. Dorokhov, B.I. Ermolaev,
S.B. Gerasimov, E.V. Shuryak  and V.Vento for useful discussions.

This work was supported in part by the Heisenberg-Landau program and
the following grants: INTAS-2000-366, RFBR-01-02-16431, RFBR-02-02-81023,
RFBR-03-02-17291.

\end{document}